\documentclass[showpacs,preprintnumbers,amsmath,amssymb]{revtex4}

\usepackage{graphicx}
\usepackage{dcolumn}
\usepackage{bm}

\usepackage{latexsym}
\usepackage{amsmath}
\usepackage{amssymb}
\usepackage{amsbsy}

\begin{document}

\title{Comment on ``Chiral Suppression of Scalar-Glueball Decay''}
\author{Z.F.Zhang and H.Y.Jin}
\affiliation{ Zhejiang Institute of Modern Physics,  Department of
Physics, Zhejiang University, Zhejiang Province, People's Republic
of China}


\maketitle

To identify glueballs in experiments is really a tough job. Based
on the perturbative theory, the author of ref.\cite{chanowitz}
suggested that the SU(3) flavor symmetry is probably strongly
broken in two-body decays of  scalar glueball. If this allegation
is valid,  mixing between scalar glueballs and $\bar qq$ quarkonia
should be chiral-suppressed. To our knowledge, such a conclusion
seems too simple and not reliable. The problem arises from the
fact that non-perturbative effects are not negligible in this
case. In the constituent quark model, the quarks dress with
gluons, so that the masses of $u$ and $d$ quarks are around 300
MeV and the mass of $s$ quark is about 400 MeV. Then the
difference between production rates of  $\bar s s$  and $\bar u
u$, $\bar dd$ should not be so large as that claimed by teh author
of reference \cite{chanowitz}. This can be directly tested by the
experimental data for the decays of $\chi_{c0}(1P)$ ($0^{++}$).
This $\bar cc$ charmonium  first decays into gluons then the
gluons convert into light hadrons. From the particle data
book\cite{pdg}, $\Gamma(\chi_{c0}(1P)\to
KK)$/$\Gamma(\chi_{c0}(1P)\to \pi\pi)$$\approx 1.1$, so the SU(3)
flavor symmetry seems to be nicely respected. Theoretically the
natural understanding of this fact is that, the U(1) chiral
symmetry can be broken by either QCD vacuum or light quark masses.
The first mechanism does not break the SU(3) flavor symmetry
whereas the second one does. The famous U(1) puzzle indicates that
the first mechanism dominates at low energy, i.e. a left-handed
fermion can convert into a right-handed  fermion by exchanging
instantons rather than through the current quark masses. Such
instanton-induced effects can be estimated in terms of the
Instanton Liquid Model. In order to make a comparison,  just
following the method given in \cite{chanowitz}, we  set the two
gluons on shell and keep only the leading order in the coupling
$\alpha_s$. Indeed, a full comparison order by order in the
coupling $\alpha_s$ expansion is impossible at present and beyond
the scope of this comment.

According Chanowitz's method\cite{chanowitz}, we obtain the ratio
of the decay widths ($\Gamma_1$ only includes the perturbative
contribution and $\Gamma_2$  refers to the instanton effects) as
\begin{equation}
\frac{\Gamma_1(G_0\rightarrow \bar ss)}{\Gamma_2(G_0 \rightarrow \bar qq)} = \frac{9}{16\pi^2} \frac{m_s^2 m^{\star 2}}{M_G^4}\frac{\ln^2 ((1+\beta)/(1-\beta))}{n^2 \rho^6 A^2 (1+\beta^2)},
\end{equation}
where $q$ can be $u$, $d$ and $s$ quarks. The parameter $A$ is
defined as
\begin{equation}
\begin{split}
A=&\int_0^{\theta_\Lambda} d \theta \sin \theta \int_0^1 d\alpha \frac{K_1 ( (M_G/2) \rho \sqrt{1+\beta^2 +2 \beta \cos \theta}/ \sqrt{\alpha(1-\alpha)})}{\alpha(1-\alpha) (M_G/2) \sqrt{1+\beta^2 +2 \beta \cos \theta}} (3+\beta \cos \theta)\\
&+\int_{\theta_\Lambda}^\pi d \theta \sin \theta \int_0^1 d\alpha \frac{K_1 ( \Lambda \rho / \sqrt{\alpha(1-\alpha)})}{\alpha(1-\alpha) (M_G/2) \sqrt{1+\beta^2 +2 \beta \cos \theta}} (3+\beta \cos \theta).
\end{split}
\end{equation}
The upper bound of the integral  $\theta_\Lambda$ is defined as
\begin{equation}
\theta_\Lambda = \cos^{-1} \frac{1}{2 \beta} (4 \Lambda^2/M_G^2 -1-\beta^2),
\end{equation}
and $\Lambda$ is a scale cut-off which removes  the infrared
divergence in the McDonald function $K_1$.

The other parameters employed in the phenomenological analysis
are\cite{shuryak, shuryak2}
\begin{gather}
\rho = 1/0.6 GeV,~~~~ n = 8 \times 10^{-4} GeV^4,\\
m^\star = 0.086 GeV,~~~~m_s=150 MeV,~~~~ M_G=1.65 GeV.
\end{gather}

The cut-off $\Lambda$ is set within the range of $200\;{\rm
MeV}\sim 500\;{\rm MeV} $. We find that $\Gamma_1(G_0 \rightarrow
\bar ss)/\Gamma_2(G_0 \rightarrow \bar qq)$ is in the region of
$0.04\sim 0.4$. It is shown that the instanton effect is  more
important. When we choose the glueball mass as $M_G=5$ GeV,
$\Gamma_1(G_0 \rightarrow \bar ss)/\Gamma_2(G_0 \rightarrow \bar
qq)$ is in the region of $0.09 \sim 1.0$. This result is
consistent with the experimental data for the two-body decays of
$\chi_{c0}(1P)$.

\begin{acknowledgments}
The work was supported by NSFC and ZJNSF.
\end{acknowledgments}

\end{document}